\documentclass{aa}
\usepackage{aabib99,epsf}
\usepackage{tabularx}

\def\infig#1#2#3{\epsfysize=#1cm \centering{\mbox{\epsfbox{#2}}}}
\newcommand{\preservebackclash}[1]{\let\temp=\\#1\let\\=\temp}

\begin{document}

\thesaurus{05(11.16.1; 11.19.1; 11.19.3; 11.19.16; 13.09.1)}

\title{Near--infrared probing of embedded structures in starburst and
  Seyfert galaxies\thanks{Based on observations collected at the
    European Southern Observatory, La Silla, Chile (ESO programme
    59.A--0773)}}

\author{
D.\ts Greusard \inst{1},
D.\ts Friedli \inst{4}, 
H.\ts Wozniak \inst{2},
L.\ts Martinet \inst{1},
P.\ts Martin \inst{3}
} 

\institute{
Observatoire de Gen\`eve, CH--1290 Sauverny, Switzerland
\and
Observatoire Astronomique de Marseille--Provence, F--13248 Marseille Cedex 4, France
\and
Canada--France--Hawaii Telescope, PO Box 1597, Kamuela, HI 96743, USA 
\and  
Gymnase de Nyon, CH-1260 Nyon, Switzerland 
}

\offprints{Didier.Greusard@obs.unige.ch}
\date{Received Mars 14, 2000; accepted June 28, 2000}

\authorrunning{Greusard et al.}
\titlerunning{NIR study of starburst and Seyfert galaxies}
\maketitle 

%_________________________________________________________________
\begin{abstract}
  Surface photometry in the J and K' bands of 15 southern Seyfert or
  starburst galaxies is presented. The detailed central morphology and
  structural properties of these objects were analyzed by fitting
  ellipses to isophotes. New central peculiar structures have been
  identified like, for instance, three double--barred systems
  (ESO\,215--G031; ESO\,320--G030; ESO\,443--G017), one object with a
  nested nuclear spiral structure at the center of a primary bar
  (NGC\,5135), one object with a nuclear bar without evidence of a
  large--scale bar (NGC\,4941), and one galaxy with a likely dissolved
  secondary bar within a primary one (ESO\,508--G005). The J--K'
  radial profile proved to be reasonably well linked with the 
  presence of a starburst, but not with the Seyfert activity. For significant
  starbursts, the central J--K' value is 0.3--1.5\,magnitude
  larger than the disc one.

\end{abstract}

%_________________________________________________________________
\section{Introduction}

\begin{table*}[t]
\caption{Sample of observed galaxies.}
\label{sample}
\begin{tabular}{llllrrrrcr} 
\hline 
%
%-> les "multicolumn{1}" sont inutiles.
%HW: ici ca a l'air de centre les legendes des colonnes tout en laissant
% les chiffres a gauche ou a droite non? Auquel cas laisser.
%
\multicolumn{1}{c}{Galaxies}       &
\multicolumn{1}{c}{$\alpha$(2000)} & 
\multicolumn{1}{c}{$\delta$(2000)} &
\multicolumn{1}{c}{Type}           &
\multicolumn{1}{c}{$z$}            &
\multicolumn{1}{c}{$D_{25}$}       &
\multicolumn{1}{c}{$i$}            &
\multicolumn{1}{c}{PA$_{\rm disc}$} & 
\multicolumn{1}{c}{Central}        &
\multicolumn{1}{c}{$t_{\rm expo}$} \\
\multicolumn{1}{c}{}               &
\multicolumn{1}{c}{}               & 
\multicolumn{1}{c}{}               &
\multicolumn{1}{c}{}     	   &
\multicolumn{1}{c}{[$10^{-3}$]}     &
\multicolumn{1}{c}{[$'$]}          &
\multicolumn{1}{c}{[$^{\circ}$]}   &
\multicolumn{1}{c}{[$^{\circ}$]}   &  
\multicolumn{1}{c}{Activity}       &
\multicolumn{1}{c}{[s]}            \\
\multicolumn{1}{c}{(1)} &
\multicolumn{1}{c}{(2)} &
\multicolumn{1}{c}{(3)} &
\multicolumn{1}{c}{(4)} & 
\multicolumn{1}{c}{(5)} &
\multicolumn{1}{c}{(6)} &
\multicolumn{1}{c}{(7)} &
\multicolumn{1}{c}{(8)} &
\multicolumn{1}{c}{(9)} & 
\multicolumn{1}{c}{(10)} \\
\hline
\object{ESO\,374--G032} 	& $10^{h}06^{m}04^{s}$	& $-33^{\circ}53'02''$ &GPair 		&34	&		&51 	&	&STB  	&1200/1200\\
\object{ESO\,264--G036} 	& $10^{h}43^{m}08^{s}$ 	& $-46^{\circ}12'39''$ &SB(s)b 		&23	&1.1$\times$0.7 &49 	&102	&   	&1200/1200\\
\object{NGC\,3393} 	& $10^{h}48^{m}24^{s}$ 	& $-25^{\circ}09'40''$ &(R')SB(s)ab 	&12	&2.2$\times$2.0	&24	&	&STB/Sy2   	&1200/1200\\
\object{ESO\,215--G031}   & $11^{h}10^{m}35^{s}$	& $-49^{\circ}06'12''$ &(R'1)SB(r)b    	&9	&2.4$\times$1.7 &41 	&130	&STB   	&1200/1200\\
\vspace{1truemm}
\object{ESO\,320--G030}	& $11^{h}53^{m}12^{s}$	& $-39^{\circ}07'49''$ &(R'1)SAB(r)a	&11	&2.2$\times$1.3	&55	&121	&STB   &1200/1200\\
\object{ESO\,443--G017}	& $12^{h}57^{m}45^{s}$	& $-29^{\circ}45'59''$ &(R)SB(r'l)0/a	&10	&1.4$\times$0.9	&50	&23	&STB	&1200/1200\\
\object{NGC\,4903}	& $13^{h}01^{m}23^{s}$	& $-30^{\circ}56'02''$ &SB(rs)c		&16	&1.5$\times$1.3	&36	&73	&Sy2   	&1200/1200\\
\object{NGC\,4941}	& $13^{h}04^{m}13^{s}$	& $-05^{\circ}33'06''$ &(R)SAB(r)ab	&4	&3.6$\times$1.9	&57	&15	&Sy2  	&1200/1200\\
\object{NGC\,4939}	& $13^{h}04^{m}14^{s}$	& $-10^{\circ}20'23''$ &SA(s)bc		&10	&5.5$\times$2.8	&59	&10	&Sy2   	&1200/1200\\
\vspace{1truemm}
\object{ESO\,323--G077}	& $13^{h}06^{m}27^{s}$	& $-40^{\circ}24'50''$ &(R)SB(l)0	&15	&1.5$\times$1.0 &47 	&155	&STB/Sy1 &1200/1200\\
\object{ESO\,508--G005}	& $13^{h}06^{m}56^{s}$	& $-23^{\circ}55'02''$ &SB(rl)0/a	&10	&1.3$\times$1.0	&41	&	&Sy2  	&1200/1200\\
\object{NGC\,5135}	& $13^{h}25^{m}44^{s}$	& $-29^{\circ}50'02''$ &SB(l)ab		&14	&2.6$\times$1.8	&45	&	&STB/Sy2   	&1200/1200\\
\object{NGC\,5643}	& $14^{h}32^{m}41^{s}$	& $-44^{\circ}10'28''$ &SAB(rs)c 	&4	&4.6$\times$4.0 &29 	&	&Sy2 	&1200/600\\
\object{NGC\,6221} 	& $16^{h}52^{m}47^{s}$	& $-59^{\circ}12'59''$ &SB(s)bc pec	&5	&3.5$\times$2.5	&46	&5	&Sy2   	&900/900\\
\object{NGC\,6300}	& $17^{h}16^{m}59^{s}$	& $-62^{\circ}59'11''$ &SB(rs)b		&4	&4.5$\times$3.0	&49	&118	&Sy2	&900/750\\
\hline
\end{tabular}
\vspace{1truemm}

Columns (2), (3), (4), (5), (6) \& (7): from the NASA/IPAC Extragalactic
Database (NED).\\
Column (8): position angle on the sky measured from North through
East; from de Vaucouleurs et al. \cite*{rc3}. \\
Column (9): Sy\,=\,Seyferts from V\'eron \& V\'eron \cite*{verver93} / 
STB\,=\,starbursts $\leftrightarrow \log(S_{60}/S_{100})\ga-0.35$ (see text). \\
Column (10): total exposure time, $x/y$ $\rightarrow$ $x$ seconds in
J--band and $y$ seconds in K'--band.
\end{table*}

Fueling of the inner region of starburst galaxies and active galactic
nuclei (AGN) is still an unresolved issue. Several clues are useful to
investigate this problem.  For example, Shlosman et
al. \cite*{shlfra_89} have suggested that bars within bars are engines
to include in mechanisms for gas fueling active nuclei. In this
scenario, the primary (large--scale) bar brings gas, through angular
momentum transfer, from the outer part of the galaxy down to few kpc
from the central region. From there, the nuclear bar could take over
and drive this gas toward the inner region. Friedli \& Martinet
\cite*{frimar93} using 3D self--consistent numerical simulations
confirmed the possibility of forming stable systems (over several Gyr)
with secondary bars within primary bars, and showed that these
features might play a key role in collecting gas into the central
region. Moreover, such embedded structures have already been observed
in several galaxies (e.g. de Vaucouleurs 1974\nocite{dev74}; Jarvis et
al. 1988\nocite{jardub_88}; Wozniak et al. 1995\nocite{wozfri_95},
Paper I).

Bringing material from the periphery to a region very close to the
central activity requires a multi--scale process, so a multi--scale
analysis appears appropriate. In addition, the fueling activity
problem seems also to depend on the activity type (AGN or starburst):
\begin{enumerate}
\item[--] {\em Primary bars (large--scale)}: contrary to other
  previous results (e.g. Ho et al. 1997\nocite{ho_fil_97}), Knapen et al.
  2000\nocite{knashl_00} suggest that Seyfert hosts are barred more often
  that non--Seyfert ones.  Starburst galaxies seem to have a higher
  incidence of primary bars (e.g. Hunt \& Malkan 1999 \nocite{hunmal99}).

\item[--] {\em Secondary bars (small--scale)}: it is not yet well
  established if they are more prevalent in Seyferts than in
  non--active galaxies (Paper I; Jungwiert et al. 1997\nocite{juncom_97})
  or not \cite{mulreg97}.
%DG: et les bar. nucl. ds les STB?
\end{enumerate}
Thus additional morphological and kinematical studies of
starburst and AGN remain necessary to give some more insight on the
host dynamics and evolution.

In view of all previous considerations, this study presents J and K'
surface photometry of fifteen starburst and Seyfert galaxies in order
to infer their morphological and structural properties. The principal
aim is to detect bars and embedded bars among objects of this sample.
The near--IR bands are less affected by dust than visual ones, and
they are well suited to study the obscured central parts of disc
galaxies. The information provided by these filters is essential to
obtain reliable parameters about the old stellar population which is
dynamically the most important one. While several near--IR imaging
studies have recently been published (Friedli et
al. 1996\nocite{friwoz_96}, Paper II ; Jungwiert et
al. 1997\nocite{juncom_97}; Mulchaey et al. 1997\nocite{mulreg_97};
Alonso--Herrero et al. 1998\nocite{alosim_98}; Peletier et
al. 1999\nocite{pelkna_99}; M\'arquez et al. 1999\nocite{mardur_99}),
the total number of galaxies observed in these wavelengths is still
modest so new data are necessary. Moreover, since galaxies might
experienced secular evolution (see Friedli 1999\nocite{fri99} for a
review), i.e. the small and large--scale morphology of galaxies could
change over cosmological time--scales ($\approx\!10\,$Gyr), it is
crucial to observe the largest sample possible of various
morphological configurations. Thus increasing appreciably near-IR data
is essential to study the birth and evolution of embedded structures.

The paper is organized as follows. The sample, observations, reduction
processes and the analysis methods are described in Sect.~2.  Sect.~3
contains the results in form of individual descriptions of the sampled
objects. The global analysis and discussions are given Sect.~4 and
our conclusions in Sect.~5.

%_________________________________________________________________
\section{Observations, data reduction and analysis}
The sample of galaxies is presented in Table~\ref{sample}. All Seyfert
2 galaxies have been selected from the V\'eron's catalog
\cite{verver93} with the following criteria: V magnitude less than 15,
not or weakly interacting, and not too inclined.  Galaxies with an
IRAS colour $\log(S_{60}/S_{100})\ga-0.35$ are considered as
starbursts (see Sect.~3.2.2) and have been picked out from
Rowan-Robinson \& Crawford \cite*{rowcra89}. Despite its
``non--activity'', ESO264--G036 has also been observed because it has
$\log(S_{60}/S_{100}) \!\approx\! -0.37$, and its
starburst component defined in Rowan--Robinson \& Crawford
\cite*{rowcra89} suggests a transition case between starburst and
non--starburst galaxies.

A visual inspection of these objects was performed to obtain the final
sample. We do not make any attempt to make our selection criteria free
of bias. For instance, our sample is obviously biased towards the
search for nuclear structure.

%___________________________
\subsection{Observations and reductions}
The observations were carried out on April 7-9, 1997, with the ESO/MPI
2.2m telescope at La Silla; the conditions were photometric with a
FWHM seeing ranging from 1.2 to 1.6$\arcsec$. The infrared imaging
camera IRAC--2b was equipped with a 256$\times$256 NICMOS--3 array
with a pixel size of 0$\arcsec$.507\,pixel$^{-1}$, giving a field of
view of 129$\arcsec$$\times$129$\arcsec$.  For each object, detector
saturation was avoided by taking a series of exposures on the object
interspersed with sky exposures. This procedure of alternating between
the source and the sky was repeating until the total integration time
in Table~\ref{sample} was reached.

Each frame was cleaned from cosmic rays as well as from cold and hot
pixels. Flat--field frames were obtained from exposures taken each
night on a uniform illuminated blank screen (dome flat--field). After
stars were removed from sky frames, the sky background was computed
using the mean of the two sky frames gathered before and after a
science frame. Galaxy frames were then sky subtracted and divided by
the normalized dome flat--field.

As they follow the rapid variation of the sky structures, mean sky
frames were preferred to median sky frames (median of all sky frames
taken over the observation of one galaxy). As they are not affected by
the intrinsic time--dependent unflatness of the sky, dome flat--fields
were preferred rather than median sky flat--fields (median of all sky
frames taken over the night).  These choices have been confirmed by
the flatness of the images, the quality of the background substraction
and the photometric calibration.

Finally, all frames of one galaxy were co--added to create a
time--cumulated science frame. As the telescope was moved between two
science expositions, galaxy frames had to be shifted with regard to
reference points; they were defined as the peak of a Gaussian adjusted
to the intensity profile of stars present in all frames of a same
object.

%___________________________
\subsection{Photometric calibrations}

%___________________________
\begin{table}[t]
\caption{Aperture photometry comparison between
	our study and published magnitudes $\Delta$J
	and $\Delta$K'.}
\label{compare}
\begin{tabular}{lcrrrr}
\hline
\multicolumn{1}{c}{Galaxies}  &
\multicolumn{1}{c}{Aperture}  &
\multicolumn{1}{c}{J$_{our}$}   &
\multicolumn{1}{c}{K$'_{our}$}  &  
\multicolumn{1}{c}{$\Delta$J} & 
\multicolumn{1}{c}{$\Delta$K'} \\ 
\multicolumn{1}{c}{}          &
\multicolumn{1}{c}{[$\arcsec$]}  &
\multicolumn{4}{c}{[mag]}     \\
\hline
NGC\,3393 (1)	&15	&11.3	&10.3	& 0.31	& 0.20 \\ 
                &30	&10.8	&9.8	& 0.31	& 0.16 \\
NGC\,4941 (2)	&14	&11.2	&10.2	&-0.21	&-0.13 \\
NGC\,5135 (3)	&12	&11.3	&10.0	&-0.07	& 0.06 \\
		&34	&10.5	&9.4	&-0.28	&-0.08 \\
NGC\,5643 (4)	&34	&10.2	&9.1	& 0.03	& 0.07 \\
		&51	&9.9	&8.8	& 0.11	& 0.20 \\
NGC\,6221 (5)	&34	&9.8	&8.7	&-0.09	&-0.03 \\
		&51	&9.4	&8.3	&-0.12	&-0.04 \\
\hline \hline
\multicolumn{4}{l}{Mean algebraic difference} &  0.00 & +0.05
\\
\multicolumn{4}{l}{Mean absolute difference} &  0.17 & 0.11
\\
\multicolumn{4}{l}{Root mean square deviation (rms)} &  0.21 & 0.12
\\
\hline
\end{tabular}
\vspace{1truemm}

(1): from Alonso--Herrero et al. \cite*{alosim_98}. \\
(2): from Dultzin--Hacyan \& Benitez \cite*{dulben94}. \\
(3), (4), (5): from Glass \& Moorwood \cite*{glamoo85}.
\end{table}

Frame calibration was achieved by observing infrared standards from
Carter and Meadows \cite*{carmea95} during each night (2 on April 7th
and 9th, and 3 on April 8th). The zero points of each night are all
compatible within their rms error. We have used the zero point
obtained from the 3 nights put together; that way the rms error in the
determination of the photometric zero points was $\approx\!0.04$ (J)
and $\approx\!0.03$ (K'). In order to check our calibration, we have
compared our results with aperture photometry taken from literature
(see Table~\ref{compare} and Fig.~\ref{figecart}). Since the
difference between K' and K is weak and subject to uncertainties
\cite{waicow92}, we have directly compared K'
with K photometry. The weak mean algebraic difference (see
Table~\ref{compare}) reveals very little systematic deviation between
our measurements and the published ones. The K' case may reflect the
slight difference between K' and K; Wainscoat \& Cowie
\cite*{waicow92} found that, for a sample of 16 A stars or M dwarfs,
the K' filter has a zero point which is 0.03--0.04\,mag fainter than
the K one. But they also emphasize that this zero point departure may
differ for a wider population of stars. Since there is very little
systematic deviation and with regard to the mean absolute differences
and rms, our calibrations are accurate enough in the context of this
morphological study.

As the airmass correction obtained from mean atmospheric extinction
coefficient for this site ($a_{J}\!=\!0.08$,
$a_{K'}\!=\!0.11$\,mag/airmass) does not improve our calibration
check, it was not applied.  No attempt was made to take the Galactic
extinction and the K--correction into account.

%___________________________
\setcounter{figure}{0}
\begin{figure}[t]
\infig{8.8}{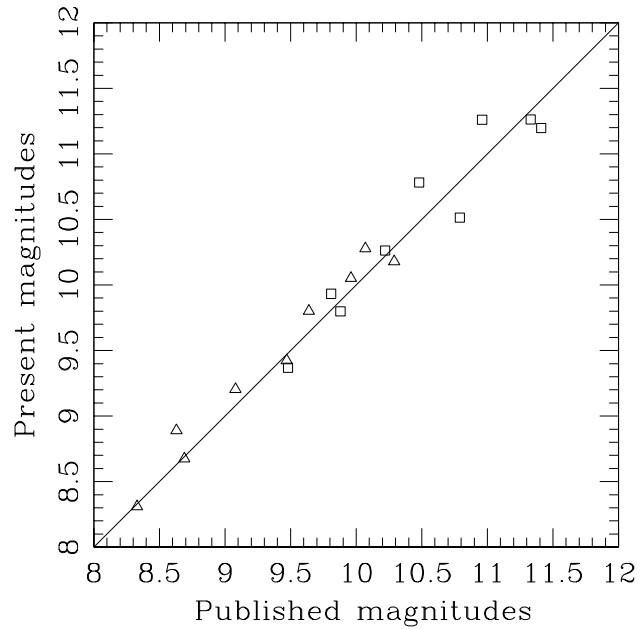}{8.8}
\caption[]{Comparison with published data of aperture photometry for 
  the objects listed in Table~\ref{compare}.  Squares are for the
  J--band and triangles for the K'--band}
\label{figecart}
\end{figure}

%_________________________________________________________________
\subsection{Surface photometry analysis}
Ellipses were fitted to isophotes of the whole sample in both J and K'
bands. As shown by various authors (e.g. Paper I), this technique
provides good qualitative and quantitative estimations of the shape of
embedded structures. As we want to observe the mean behaviour of
isophotes, field stars and regions of intense emission (e.g. giant
star forming regions) were flagged on galaxy frames before ellipse
fitting.  The program used is described in Paper I (and references
therein).  Each ellipse is characterized by:
\begin{itemize}
\item its semi--major axis length, $a$ [$\arcsec$];
\item its ellipticity $e$, $e\!=\!1-b/a$ where $b$ is the semi--minor
  axis;
\item its intensity converted into surface brightness $\mu$,
  [mag\,$\arcsec$$^{-2}$];
\item its position angle, PA [$^{\circ}$], which is the angle between
  the semi--major axis and the North. This angle was defined to match
  the conventional notation: North is up, East at the left. Angles are
  measured from North and are positive in the anti--clockwise direction
  modulo 180$^{\circ}$.
\end{itemize}
Thus, in each band, $\mu (a)$, $e(a)$, and PA$(a)$ profiles are
derived, with $a$ increasing by a factor 1.01 between each point. In
addition, ``differential profiles'' ($\mu_{J}-\mu_{K'}$,
$e_{J}-e_{K'}$, and PA$_{J}-$PA$_{K'}$) was computed as the difference
between profiles in each band at the same semi--major axis. For that
purpose individual profiles were linearly interpolate between each
point.

%_________________________________________________________________
\subsection{HST and literature data}
Because of their high spatial resolution, HST frames are useful to
visually detect and/or confirm the presence of potential central
asymmetries. Thus, when available, WFPC2 and/or NICMOS calibrated
frames have been obtained from the Hubble Data Archive. The planetary
camera CCD of WFPC2 instrument generally gave images of the galaxy
centers in the F606W filter, with a pixel size 0$\arcsec$.0455 and a
field of view of 37$\arcsec$x37$\arcsec$. NICMOS frames were gathered
with the F160W filter ($ \!\approx $H band) and have a pixel size of
0$\arcsec$.075 /pixel and a field of view of
19$\arcsec$.2x19$\arcsec$.2.

In Sect.~3.2.2, where the behaviour of the NIR colour profile versus
IRAS colour is studied, data from the literature have enriched our
sample. Galaxies for which photometric J, K' frames and IRAS data were
available were found in Alonso--Herrero et al. \cite*{alosim_98} and in
Paper II. Hence J and K' frames (with different pixel sizes
0$\arcsec$.286\,pixel$^{-1}$ or 0$\arcsec$.143\,pixel$^{-1}$) of 7
galaxies were obtained from the former study, as well as J and K
frames (with different pixel sizes 0$\arcsec$.49\,pixel$^{-1}$,
0$\arcsec$.6\,pixel$^{-1}$, or 0$\arcsec$.9\,pixel$^{-1}$) of 5
galaxies from the latter.  Ellipse fitting was performed on those
galaxies in exactly the same way as for our sample.

%___________________________
\subsection{Terminology}
We use in this paper the following terminology:
\begin{itemize}
\item {\em Primary bar:} a bar with a scale--length of several kpc.
\item {\em Secondary bar:} a bar with a scale--length around one
  kpc or less, and nested within a primary bar.
\item {\em Nuclear bar:} the same feature as the secondary bar but with no
  evidence for a primary bar.
\item $a_p$ ($a_s$), $e^{\rm max}_p$ ($e^{\rm max}_s$), PA$_p$
  (PA$_s$) are respectively the length, the maximum ellipticity, and
  the position angle of the primary (secondary) bar.
\item $\theta_{12}$ is the angle between the two bars. Positive values
  are for leading secondary bars whereas negative values are for
  trailing secondary bars (with respect to the primary bar).
\item $\beta_{12} \equiv a_p/a_s$.
\item $\gamma_{12} \equiv L_{p}/L_{s}$ where $L_{p} {\rm\ and\ } L_{s}$
  are the luminosity of the primary and secondary bars.
\end{itemize}

All the definitions above but the nuclear bar follow those of Paper I
and II, and all the parameters have been computed in the same way as
in Paper I.

%_________________________________________________________________
\section{Results}
The set of Figs.~\ref{figprofil} presents for each object in J and
K'--bands:
\begin{itemize}
\item the 2D image;
\item the 1D profiles of surface brightness, ellipticity, and position
angle, as a function of the semi--major axis length;
\item the J--K' colour map;
\item the differential profiles of surface brightness
($\mu_{J}-\mu_{K'}$), ellipticity ($e_{J}-e_{K'}$), and position angle
(PA$_{J}-$PA$_{K'}$).
\end{itemize}

In addition, when available, HST NICMOS images are also shown in
Figs.~\ref{figprofil}.
 
We hereafter describe the outstanding features and structural
parameters of each galaxy, as they have been inferred from
Figs.~\ref{figprofil}, HST images and from the literature.  This
presentation is not exhaustive; only the characteristics important in
the context of this paper are emphasized. All quoted value of the
structural parameters refer to the K'--band unless otherwise
specified.  These parameters are summarized in Table~\ref{param} for
galaxies with embedded structures. All linear distances have been
computed assuming $H_{0}\!=\!75\,{\rm km\,s}^{-1}\,{\rm Mpc}^{-1}$.

%_______________________________________
\subsection{ESO\,374--G032 (Gpair, STB, $1\arcsec\!\approx\!0.66$\,kpc)} %Interaction
The morphology of this galaxy shows evidence of recent or ongoing
interaction (mainly a large tidal tail and a disturbed central
feature). From the original IR frames, ESO\,374--G032 shows at least
two bright components within the central region separated by around
4$\arcsec$ ($\approx\! 2.6\,$kpc). This complex central structure,
also appearing in the visible frame of Kazès et al. \cite*{kazpro_90},
may be a late stage merger. This central feature results in a
bar--like contour map and ellipticity profiles. But it is interesting
to note that the PA does not show any plateau in the region of this
bar. Indeed, it decreases almost linearly from the centre up to
$\approx\!  17\arcsec$.  This characteristic could be the signature of
the strong interaction, the bar being strongly triaxial and perhaps
not coplanar.

\subsection{ESO\,264--G036 (SBb, $1\arcsec\!\approx\!0.45$\,kpc)} %B
This is a single--barred object. The well--defined primary bar ends at
$\approx\!18\arcsec$ where a slight twist is observed due to the
start of the spiral arms. In the inner region ($a \leq 2.5\arcsec$),
the $\mu_{J}-\mu_{K'}$ profile steeply decreases from $\approx\!
1.5$\,mag\,$\arcsec^{-2}$ at the center to
$\approx\!1.1$\,mag\,$\arcsec^{-2}$. The $e_{J}-e_{K'}$ profile is
roughly constant at $\approx\! 0.03$, while the PA$_{J}-$PA$_{K'}$
profile increases from about $-10^{\circ}$ to $\approx\!
5^{\circ}$. For $a \geq 2.5\arcsec$, the $\mu_{J}-\mu_{K'}$ profile
slowly increases again up to
$\approx\!1.4$\,mag\,$\arcsec^{-2}$. This feature corresponds to two
red and elongated features, probably star--forming region, close to
the bar ends.
\subsection{NGC\,3393 (SBab, STB/Sy2, $1\arcsec\!\approx\!0.23$\,kpc)} %B+B
We confirm previous observations by Jungwiert et al. \cite*{juncom_97} and
Alonso--Herrero et al. \cite*{alosim_98} that this is a double--barred
galaxy.  In close agreement with these authors, the primary bar is
oriented at PA$_p \!\approx\! 159^{\circ}$, whereas the trailing
secondary bar has PA$_s \!\approx\! 145^{\circ}$.

Differential profiles and colour plate show unusual structures:
surprisingly, the secondary bar is more prominent on J profiles than
on K' ones, leading to a small bump in the $e_{J}-e_{K'}$ profile. The
$\mu_{J}-\mu_{K'}$ profile presents an artificial negative gradient in
the inner region due to the seeing departure from the two IR bands;
this gradient must be an artefact because field stars also have a
central blue dip in the J--K' colour frame. The same features are
reproduced for other objects, when the seeing is artificially degraded
in one of the two bands. In Sect.~3.2.2, we use J and K--frames from
Alonso--Herrero et al. \cite*{alosim_98} to avoid this problem.

\subsection{ESO\,215--G031 (SBb, STB, $1\arcsec\!\approx\!0.17$\,kpc)} %B+B
This SBb starburst galaxy can be considered as the prototype of nearly
aligned double--barred systems: the ellipticity profile clearly shows
two bumps along which the position angle is nearly constant in both
bands. Those bars also tend to reduce the slope of the $\mu_{J}$ and
$\mu_{K'}$ profiles along each bar. One of the most striking feature
of this object is that the two bars are close to be aligned, $\rm PA_p
\!\approx\! 147^\circ$ whereas $\rm PA_s \!\approx\! 153^\circ$,
leading to $\theta_{12} \!\approx\!  6^{\circ}$. The ellipticity between
those bars declines close to zero at $a \!\approx\! 9\arcsec$, where
the PA variation reveals the presence of a nuclear ring. It is
observable on the J--K' image as well. As this galaxy has both inner
and outer rings \cite{but95}, this is a three--ringed galaxy.

Three distinct regions of redder J--K' colour ($ \!\approx\!
1.3$\,mag\,$\arcsec^{-2}$) are visible on this structure. Two
``hotspots'' or ``twin--peaks'' are on the ring and are symmetric with
respect to the center, the third one being at the galaxy nucleus. The
$\mu_{J}-\mu_{K'}$ profile is nearly constant at
1.1\,mag\,$\arcsec^{-2}$ along the secondary bar and at
1\,mag\,$\arcsec^{-2}$ beyond. The $e_{J}-e_{K'}$ profile shows a
slight negative gradient along the secondary bar, and is close to zero
elsewhere.
%?? The PA$_{J}-$PA$_{K'}$ profile ?
%DIDIER: COMPARER AVEC M100 PEUT-ETRE (KNAPEN ET AL. 1995, APJ 443, L73)

\subsection{ESO\,320--G030 (SABa, STB, $1\arcsec\!\approx\!0.21$\,kpc)}
%($1\arcsec \!\approx \! 213$\,pc)} %B+B
%
%DIDIER: EN FAIT, JE ME DEMANDE S'IL NE FAUDRAIT PAS LA METTRE EN B+T
%        CAR e_s^{max} EST SITUE DANS UNE ZONE OU PA_s VARIE BEAUCOUP?
The visual aspect, confirmed by the profile examination, leads us to
classify this galaxy as a new double--barred one. The primary bar has a
PA$_p \!\approx\!142^{\circ}$, different from the
PA$_{disc}=121^{\circ}$ (see Table~1), so that this objet should be classified as an
SB rather than SAB. This bar also presents two regions of slightly higher
surface brightness at its ends.
%HW , which is likely due to the shape of
%HW $x_1$ orbit family as for the case of NGC\,7020??** (Wozniak et al.
%HW \cite{woz}).  
% POUR 7020 CA NE PEUT PAS ETRE LES X1 CA RESSEMBLE PLUTOT A DES BRAS
% SPIRAUX VU DANS UNE GALAXIE TRES INCLINEE. C'EST L'ACCUMULATION SUR
% LA LIGNE DE VISEE DE LA LUMIERE A LA COURBURE DE LA SPIRALE
% POUR CET OBJET JE SERAI UN PEU DU MEME AVIS ET DONC LES DEUX SPOTS
% NON RIEN A VOIR AVEC UN BOUT DE BARRE OU SINON CELA VEUT DIRE QUE
% CETTE GALAXIE N'A PAS DE DISQUE.
% JE CLASSERAI PLUTOT EN BARRE NUCLEAIRE AVEC BULBE TRIAXIAL AUTOUR.
The secondary bar has its outermost isophotes twisted, a nearly
constant ellipticity, and PA$_s \!\approx\! 107^{\circ}$ (i.e.
$\theta_{12} \!\approx\!35^{\circ}$).

An extended central structure, clearly visible on the J--K' colour
map, is aligned with the primary bar. This might be a nuclear ring or
disc.  Nested at the center of this feature, a region of redder colour
($ \!\approx\! 1.4$\,mag\,$\arcsec^{-2}$) likely corresponds to the
secondary bar, but the angular resolution is too poor to reveal
unambiguously the bar morphology.  The $\mu_{J}-\mu_{K'}$ profile
decreases from $ \!\approx\! 1.4$\,mag\,$\arcsec^{-2}$ to reach a
constant value ($ \!\approx\!0.9$) beyond $a \!\approx\!
9\arcsec$. The $e_{J}-e_{K'}$ and PA$_{J}-$PA$_{K'}$ profiles are
roughly zero throughout the galaxy, except for the inner region where
they amount to $ \!\approx\! 0.04$ and $ \!\approx\! 10^{\circ}$
respectively.

\subsection{ESO\,443--G017 (SB0/a, STB, $1\arcsec\!\approx\!0.19$\,kpc)} %B+B
This galaxy presents a primary bar with PA$_p \!\approx\! 12^{\circ}$
that is twisted near its ends as the spiral arms begin to affect the
ellipse fitting. The small bump in ellipticity and the constant PA$_s
\!\approx\! 31^\circ$ suggest the presence of a weak secondary
bar. This secondary bar is more pronounced in K' profiles than in J
ones, resulting in a $e_{J}-e_{K'}$ excess in the central region; it
is also visible on the J--K' colour map. However, the disc inclination
and PA (Table~\ref{sample}) prevent any reliable detection from being
made. A high inclination angle make it difficult to distinguish
between a bar and a projected triaxial structure having the same PA as
the disc. The projected angle between both bars would be $\theta_{12}
\!\approx\! -19^{\circ}$. The $\mu_{J}-\mu_{K'}$ profile is nearly
constant ($ \!\approx\!  1.3$\,mag\,arcsec$^{-2}$) along the secondary
bar, and then decreases down to a constant value of $ \!\approx\!
1$\,mag\,arcsec$^{-2}$ at larger distances.

\subsection{NGC\,4903  (SBc, Sy2, $1\arcsec\!\approx\!0.31$\,kpc)} %B+L?
A very strong ($e^{max} \!\approx\! 0.7$) and quite long ($a_{p}
\!\approx\! 21\,\arcsec$) primary bar is visible.  It is surrounded by
a weakly barred component whose extent and surface brightness might
suggest a lens nature. The central part of this galaxy seems to be
relatively featureless, even if the very center shows a constant
ellipticity ($e \!\approx\! 0.1$) accompanied by a PA variation, in
contrast with the profile behaviour at larger radii. This is
reinforced by the $\mu_{J}-\mu_{K'}$ gradient observed in the same
region, which is also visible on the J--K' frame. But this region
needs a higher angular resolution to be reliably probed.
%_______________________________________
\begin{table*}[t]
\caption{Non--deprojected morphological parameters in the K'--band for galaxies
  with embedded structures.}
\label{param}
\begin{flushleft}
\begin{tabular}{lcrrrrrrrrr}
\hline
\multicolumn{1}{c}{Galaxies}        &
\multicolumn{1}{c}{Type}            &
\multicolumn{3}{c}{1st component}   &
\multicolumn{3}{c}{2nd component}   &
\multicolumn{3}{c}{Ratios}          \\
\multicolumn{2}{c}{}                &
\multicolumn{1}{c}{$a_1$}           &
\multicolumn{1}{c}{$e_1^{\rm max}$} &  
\multicolumn{1}{c}{PA$_1$}          & 
\multicolumn{1}{c}{$a_2$}           &
\multicolumn{1}{c}{$e_2^{\rm max}$} &  
\multicolumn{1}{c}{PA$_2$}          &
\multicolumn{1}{c}{$\beta_{12}$}    & 
\multicolumn{1}{c}{$\theta_{12}$}   &
\multicolumn{1}{c}{$\gamma_{12}$}\\ 
 & &
\multicolumn{1}{c}{$[\arcsec]$}           &
\multicolumn{1}{c}{} &  
\multicolumn{1}{c}{$[^{\circ}]$}    & 
\multicolumn{1}{c}{$[\arcsec]$}           &
\multicolumn{1}{c}{} &  
\multicolumn{1}{c}{$[^{\circ}]$}          &
\multicolumn{1}{c}{}    & 
\multicolumn{1}{c}{$[^{\circ}]$}   &
\multicolumn{1}{c}{}\\ 

\hline
NGC\,3393 & B+B & 19.7 & 0.44 & 159 & 3.5 & 0.12 & 145 & 5.6 & --14&2.4 \\
%\vspace{1truemm}
ESO\,215--G031 & B+B  & $\!\approx\! 47$ & 0.63 & 147 & 9.5 & 0.48 & 153 & $\!\approx\!4.9$ &   6&1.7\\
%\vspace{1truemm}
ESO\,320--G030 & B+B & $\!\approx\! 37$ & 0.64 & 142 & 5.2 & 0.32 & 107 & $\!\approx\!7.1$ &  35&2.1\\
\vspace{1truemm}
ESO\,443--G017 & B+B  & 14.7 & 0.48 &  12 & 5.5 & 0.28 &  31 & 2.7 & --19&1.5 \\
%\vspace{1truemm}
ESO\,323--G077 & B+T & 17.5 & 0.25 & 56 & 7.7 & 0.18 & 156 & 2.3 & 100&1.2 \\
%\vspace{1truemm}
ESO\,508--G005 & B+dB? & 30.9 & 0.53 & 161 & 3.6 & 0.10 &  35 & 8.6 & --54&1.8 \\ 
%\vspace{1truemm}
NGC\,5135     & B+nS & 46.1 & 0.63 & 127 & -- & -- & -- & -- & -- & --\\
%\vspace{1truemm}
NGC\,6221     & T+B & $\!\approx\!29$ & 0.53 & 116 & 5.7 & 0.29 & 5 & $\!\approx\!5.1$& --69&3.0\\
\hline
\end{tabular}
\end{flushleft}
$\approx$ $\leftrightarrow$ eye--based estimation.\\
X+Y $\leftrightarrow$ X qualifies large--scale
structures (1st component), whereas Y stands for small scale ones (2nd
component);\\
B$\,=\,$Barred isophotes, T$\,=\,$Twisted
  isophotes, nS$\,=\,$nuclear Spiral--like feature, dB$\,=\,$``dissolved Bar''.
\end{table*}

\subsection{NGC\,4941 (SABab, Sy2, $1\arcsec\!\approx\!0.08$\,kpc)} %nB
The $\mu$, $e$, and PA profiles all indicate the presence of a nuclear
bar, without any large--scale bar. The nuclear bar, clearly visible on
the NICMOS F160W frame, is characterized by $\rm PA_{s} \!\approx\!
180^{\circ}$, $e_{s}^{\rm max} \!\approx\! 0.32$, and $a_{s}
\!\approx\! 5\,\arcsec$. NGC\,7702, which also harbour a nuclear bar
($e_{s}^{\rm max} \!\approx\! 0.43 $, $a_{s} \!\approx\! 10\,\arcsec$,
see Paper I), is visually very close to this galaxy; even the
orientation of the bar relative to the disc major axis are similar
($|{\rm PA}_{s}-{\rm PA}_{disc}|\approx 20^{\circ}$).

The $\mu_{J}-\mu_{K'}$ profile shows a slight decrease along the
nuclear bar and is nearly constant afterwards ($ \!\approx\!
1$\,mag\,$\arcsec^{-2}$).

\subsection{NGC\,4939 (SAbc, Sy2, $1\arcsec\!\approx\!0.19$\,kpc)} %unbarred
The regular decrease of the surface brightness, the roughly constant
PA at the same angle as the disc ($ \!\approx\! 10^{\circ}$, see
Table~\ref{sample}), and the high inclination angle of this galaxy,
all indicate that the ellipticity behaviour should mostly be due to
projection effects of a disc plus spiral arms (see e.g. Jungwiert
et al. 1997 \nocite{juncom_97} for examples of projection effects). Thus, this
galaxy really seems unbarred, even if the central isophotes enhanced in the
NICMOS F160W frame, have a boxy shape.

Most striking is the big difference in ellipticity between J and
K' band in the central region ($\mu_{J}-\mu_{K'} \!\leq\! 0.1$ at
the center).

\subsection{ESO\,323--G077 (SB0, STB/Sy1, $1\arcsec\!\approx\!0.29$\,kpc)} %B+T
%($1\arcsec \!\approx \! 291$\,pc)}
%DIDIER: IL FAUT QUE LE PIC SOIT VISIBLE SUR LA FIGURE J-K
%        DIRE UN MOT SUR LA DIFFERENCE DES e; TROISIEME BARRE EN K?
The relatively weak primary bar is oriented at PA$_p \!\approx\!
56^\circ$, and ends near $ \!\approx\! 17.5\,\arcsec$. The inner
prominent structure has a moderate ellipticity, and presents a
monotonous PA increase from about 150$^\circ$ to 175$^\circ$, close to
the PA value of the disc (Table~\ref{sample}). These isophote twists
certainly indicate the presence of a triaxial bulge. The
$\mu_{J}-\mu_{K'}$ profile reaches the reddest colour of our sample at
the center of the galaxy ($ \!\approx\! 2.2$\,mag\,$\arcsec^{-2}$).
This extremely red, roughly axisymmetric region, is striking on the
J--K' map. Moreover all the differential profiles show strong
departure from zero toward the center.
%?? citer mulchaey 96 OIII,NII,Halpha

\subsection{ESO\,508--G005 (SB0/a, Sy2, $1\arcsec\!\approx\!0.19$\,kpc)} %dB+B
This galaxy harbours a clear primary bar with PA$_p \!\approx\!
161^\circ$.  The central region ($a\!\leq\!4\,\arcsec$) presents a
nearly constant and low ellipticity ($e\!\approx\!0.1$) with a twisted
PA around 35$^\circ$. These profiles clearly suggest the presence of a
nuclear component (as to be compared with the profile of a ``pure''
single--barred galaxy like e.g.  ESO\,264--G036). Moreover, this region
is also slightly redder: the $\mu_{J}-\mu_{K'}$ profile displays a
central plateau at $ \!\approx\! 1.1$\,mag\,$\arcsec^{-2}$, whereas
beyond another plateau has $ \!\approx\! 0.9$\,mag\,$\arcsec^{-2}$.  A
similar feature is observed in the double--barred galaxy
ESO\,215--G031 and ESO\,443--G017!

As PA$_{s}$ fairly differs from PA$_{disc}$, the central morphology is
unlikely to result from the projection of a triaxial bulge. Therefore,
an appealing and plausible possibility might be that this component is
simply the ``remnant'' of a secondary bar, or a secondary bar in the
process of dissolution (for details see
e.g. Friedli~1999\nocite{fri99}). This hypothesis might for instance be
checked via stellar and gaseous kinematics to detect fossil
signatures of previous potential asymmetries.

\subsection{NGC\,5135 (SBab, STB/Sy2, $1\arcsec\!\approx\!0.27$\,kpc)} %B+T+B or nsa+B or B+nsa+B?
%DIDIER: ICI UTILE DE VERIFIER AVEC HST, COMPARER LES PROFILES H(HST) ET K'
%        EVENTUELLEMENT V-H(HST)
The most striking feature of this galaxy is its very special center.
In view of the $\mu$, $e$, and PA profiles, one could think of a
three--component object, i.e. both primary (PA $\approx\! 130^\circ$)
and secondary (PA $\approx\! 140^\circ$) bars separated by a triaxial
bulge.  However, the visual inspection of J, K' frames instead
suggests the presence of a nuclear bi--symmetric spiral structure
nested in the primary bar. This inner component is highlighted on the
J--K' colour map. It is also visible on the WFPC2 image
\cite{malgor_98} as a dusty and patchy flocculent nuclear spiral.  The
peculiar nature of this nucleus is strengthened by the highly
disturbed differential profiles, which reveal the very unusual
differences between J and K'--bands.
%DIDIER: EVENTUELLEMENT MENTIONNER LAINE ET AL. 1999, MNRAS, IN PRESS
%        QUI TROUVENT AUSSI DES SPIRALES NUCLEAIRES DANS NGC5248.

Even on high resolution HST frames, it is very difficult to
know if these spiral arms really start from a nuclear bar. Indeed, due
to the high amounts of dust and the intense star forming regions,
ellipse fits are problematical even on NICMOS F160W images.

\subsection{NGC\,5643 (SABc, Sy2, $1\arcsec\!\approx\!0.08$\,kpc)} %B
%Single-barred galaxy.  J--K' has a GF profile.  In the flat part, the
%J--K' value is relatively high ($\approx$1.2--1.3).
%??HST, Jungwiert
Both NIR frames shows abnormal structures in the southern region;
reflection on a misaligned filter could give rise to such ghosts (none
of these structures are visible on the DSS optical frame). So the
colour gradient in the J--K' colour frame is an artefact. However, as
the central region of interest is less affected, ellipse fitting was
performed on this area. Following its NIR morphology, we suggest to
classify this galaxy as an SB one rather than SAB. The
$\mu_{J}-\mu_{K'}$ profile is nearly constant $ \!\approx\!
1$\,mag\,$\arcsec^{-2}$ and present a slight increase near the center
($a\!\leq\!1.1\,\arcsec$).

\subsection{NGC\, 6221 (SBbc pec, Sy2, $1\arcsec\!\approx\!0.10$\,kpc)} %B+T
%Nuclear bar plus a twisted, probably forming, primary bar.
%Spectacular southern spiral arm.  Significant negative gradient in the
%J--K' profile (from $\approx$1.4 to $\approx$0.7).
%Both near-IR profiles clearly confirm the presence of a nuclear bar,
%as suggested by Koribalski (\cite{kori}).
This galaxy may be in weak interaction with NGC\,6215 \cite{kor96a},
and with two low--surface brightness galaxies \cite{kor96b}. The
classification of this galaxy is somewhat controversial since de
Vaucouleurs et al. \cite*{rc3} classifies it as a barred spiral, SBc,
whereas Sandage \& Tammann \cite*{santam81} classifies it as an
ordinary spiral, Sbc. This uncertainty probably comes from the dual
nature of this object.  Because it exhibits properties of both early
and late--type, Vega Beltr\'an et al. \cite*{vegzei_98} state that
NGC\,6221 is an intermediate snapshot of a spiral evolving from early
to late--type. Nevertheless near--IR profiles revealed a clear nuclear
bar (already mentioned in Forbes \& Norris 1998\nocite{fornor98}) nested in
a twisted structure. N--body simulations \cite{pfe99} have
shown that such an object could evolve toward a double--barred system
as the shape of the large--scale twisted isophotes becoming more and
more bar--like after several dynamical time--scale.

Tsvetanov \& Petrosian \cite*{tsvpet95} have detected 173
\ion{H}{ii} regions throughout the galaxy. The more intensive ones are
mostly located along the primary bar, near the center and along the
spiral arms, and are also traced by the big redder structures visible
on our J--K' colour plate. Thus even if the IRAS indicator does not
lead us to classify this object as starburst ($\log(S_{60}/S_{100})
\!\approx\! -0.36$), it experiences active star formation.

Vega Beltr\'an et al. \cite*{vegzei_98} also detect a ring--like structure
of ionized gas with a radius of about 9$\arcsec$. Moreover this ring
is related to the presence of an inner Lindblad resonance (ILR) of the
primary bar.  At this radius the ellipticity effectively reach a
minimum, but no ring is clearly visible on the J--K' colour plate.

This galaxy has a $\mu_{J}-\mu_{K'}$ profile similar to that of
previous objects like ESO\,443--G017.

\subsection{NGC\,6300 (SBb, Sy2, $1\arcsec\!\approx\!0.08$\,kpc)} %B?
%J--K' has a GF profile.  J--K' values are high, both at the center
%($\approx$2, the highest of the sample) and in the flat part
%($\approx$1.2--1.3).
%
% le PA de Mulchaey est faux: inversion est-ouest sur leur image
Already outlined by Mulchaey et al. \cite*{mulreg_97}, NGC\,6300 harbors a
primary bar ($a_p\!\approx\!50\,\arcsec$), whereas we could not
confirm the presence of the secondary bar they also suspected.
Although this object experiences active star formation like NGC\,6221,
no strong starburst seems to be active near the central region, except
for intense nuclear emission (Crocker et al. 1996\nocite{crobau_96} report
that \ion{H}{ii} regions are mostly concentrated in the inner ring at
$a\!\approx\! 42\arcsec$). The $\mu_{J}-\mu_{K'}$ profile is roughly
constant outside the center and is steeply increasing inside
$a\!\la\!1.2\,\arcsec$.

%_______________________________________
\section{Global properties and discussion}
\subsection{Near--IR morphologies}
The structural parameters of the 8 galaxies with embedded structures
are summarized in Table~\ref{param}. Obviously no reliable statistics
could be performed on such a restricted and biased sample, but it enriches
previous studies. Note that morphological parameters are difficult to
compare with those of the literature since each author uses various
definitions and computation methods. Thus we will make a comparison only
with Paper II.

For the B+B category (NGC\,3393, ESO\,215--G031, ESO\,320--G030,
ESO\,443--G017), ranges of bar lengths and luminosity ratios
($2.7\!\leq\!\beta_{12}\!\leq\!7.1$,
$1.5\!\leq\!\gamma_{12}\!\leq\!2.4$ ) are shifted toward lower value
than in Paper II.
($4.0\!\leq\!\beta_{12}\!\leq\!13.4$,
$1.8\!\leq\!\gamma_{12}\!\leq\!7.5$). Thus our new data tend to
decrease the mean $\beta_{12}$ and $\gamma_{12}$ from $7.2$ and $3.6$
down to $6.5$ and $3.0$ respectively. Those four double--barred
galaxies confirm that no preferential angle are found between the two
bars, so that nested bars are dynamically decoupled.  Moreover while
there is a lack of known objects with low $\theta_{12}$, two galaxies
have nearly aligned bars ($|\theta_{12}|\!\approx\!6^{\circ}$ and
$|\theta_{12}|\!\approx\!14^{\circ}$, in ESO\,215--G031 and NGC\,3393
respectively).

Some galaxies (ESO\,264--G036, NGC\,3393, NGC\,4939, ESO\,323--G077,
NGC\,5135, NGC\, 6221) exhibits significant differences in their
isophote shapes (PA and $e$) between the two NIR bands. Quillen et al.
\cite*{quiram_96} have suggested that non--circular stellar motion and
radial colour gradient (resulting from the stellar population
gradient), could lead to such ellipticity differences between the two
bands. But, in the present case, the ellipticity deviation occurs in
the central region where some of the Quillen et al. assumptions are
not respected; principally the colour gradient could be strongly
affected by dust lanes and/or star formation. Moreover, as we have
seen in the case of NGC\,3393, isophote shape deviations on
small--scale could also result from seeing discrepancy between the two
bands.
%        TYPE DE HUBBLE et en L_FIR.

\subsection{Behaviour of $\mu_{J}-\mu_{K'}$ profiles}
%___________________________
\setcounter{figure}{2}
\begin{figure}[t]
%\resizebox{8.8cm}{!}{\includegraphics{ds1865_fig2.ps}}
\infig{8.8}{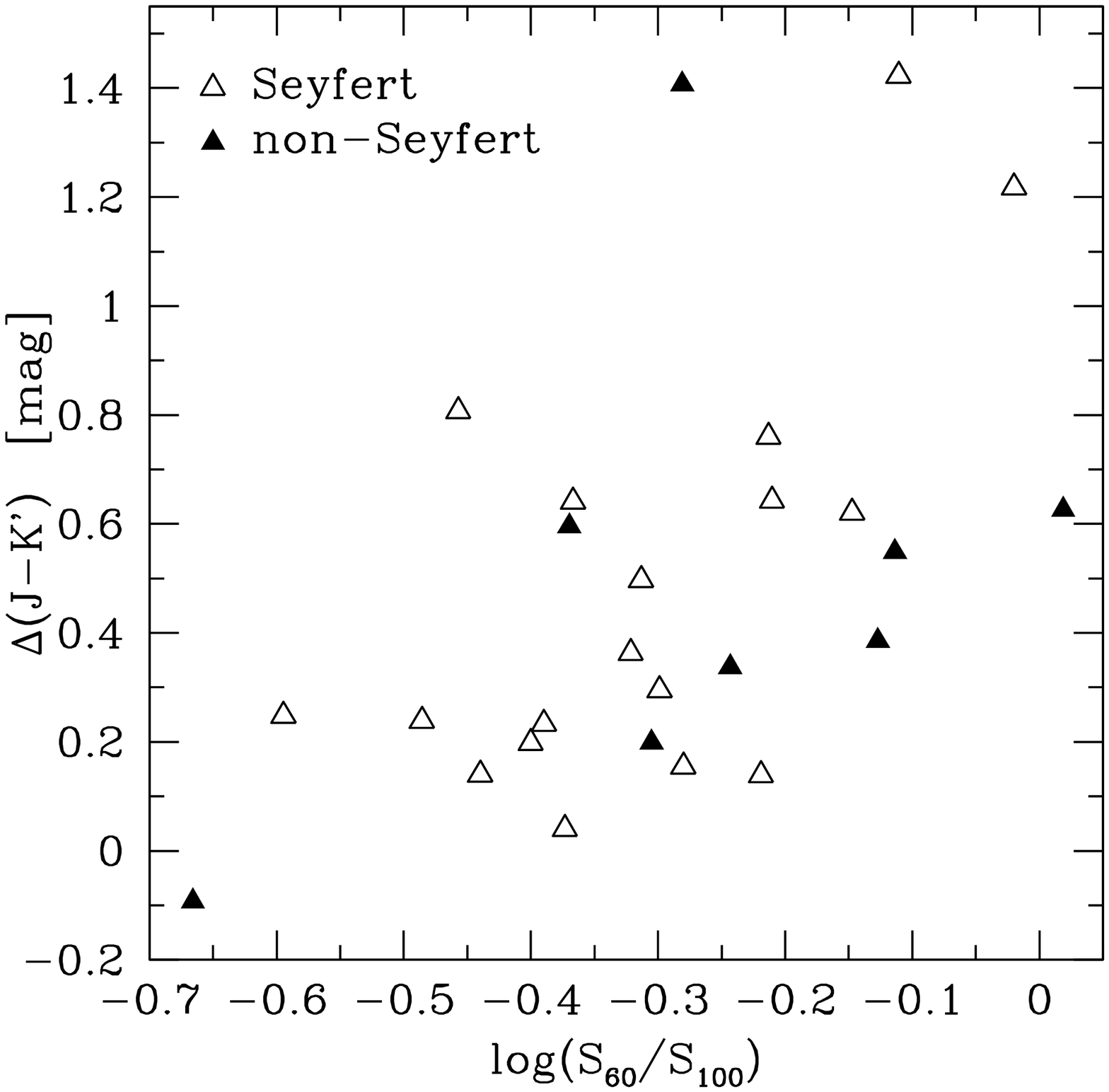}{8.8}
\caption[]{Central $\mu_{J}-\mu_{K'}$ profile behaviour versus star formation
intensity (see text). The whole sample is presented in Table~\ref{djmktab}. Seyfert stands for both Seyfert
and/or Liner objects.
% Nuclear starburst galaxies tend to be located in the upper right corner
}
\label{fig3}
\end{figure}
%___________________________
Whereas all $\mu_{J}-\mu_{K'}$ profiles are roughly constant outside
the inner region ($\mu_{J}-\mu_{K'}\!\approx\!1\,{\rm
mag}\arcsec^{-2}$ at R$\la$4--5\arcsec), all of them increases toward
the center or at least are indicative of redder central J--K' colour
(if we exclude NGC\,3393, see Sect.~3.1.3).  These $\mu_{J}-\mu_{K'}$
profiles are qualitatively and quantitatively very similar to those
published in Hunt et al. \cite*{hunmal_97}.

Could the central activities give rise to such central NIR colour
profile features? For studying this issue, we have reduced the
$\mu_{J}-\mu_{K'}$ profile behaviour to one single parameter
$\Delta$(J--K'). It is the difference between the inner and the outer
J--K' colour. This differential colour has the definite advantage of
being independent of the photometric calibration and corrections; in
particular it is independent of the airmass, the Galactic extinction
or the K correction. The inner J--K' colour is defined to be the
difference between $\mu_{J}$ and $\mu_{K'}$, both integrated within
the fitted ellipse $a_{0}\,=\,$1.5$\arcsec$ semi--major axis. The
outer J--K' colour is computed between the semi--major axis $a_{1}$
and $1.2\,a_{1}$, where $a_{1}\,=\,0.1D_{25}$ and $D_{25}$ is taken
from de Vaucouleurs et al.  (1993). Thus this outer aperture has a
linear spatial extent roughly equivalent for all bright galaxies,
regardless to its distance.  Concerning the inner NIR colour, $a_{0}$
has to be as short as possible in order to probe a region affected as
little as possible by activities other than the nuclear ones
(e.g. rings of star forming regions). Ideally $a_{0}$ should also be
scaled to $D_{25}$ to make it consistent for all our sample, but we
consider that the smallest significant aperture must at least have two
pixels width ($\sim a_{0}$=1\,pixel).  $a_{0}\,=\,$1.5$\arcsec$
respects this requirement and avoids, for our sample, the problem of
contamination by nuclear rings. However, it must be used carefully for
more distant objects.

If the central profile behaviour is linked to the Seyfert activity,
$\Delta$(J--K') should be different for Seyfert and non--Seyfert
population. The hypothetic link with starburst activity is less
straightforward to enlighten since all disc galaxies are forming stars
(and not only in the central region). As the starburst classification is
almost arbitrary, we prefer to study the behaviour of $\Delta$(J--K')
as a fonction of the star formation
rate (SFR). The choice of a SFR estimator is dicussed below.

As the dust absorption peaks in the UV, the dust (mainly the big
grains component of D\'esert et al. 1990\nocite{desbou_90} model) in
thermal equilibrium with the radiation field reaches a temperature
which mostly depends on the UV--flux (in fact regardless to the nature
of the heating sources). Thus the dust temperature is well suited to
study the intensity of the activity which dominates the
UV--emission. It also has the advantage of probing these source
intensities without requiring spectroscopic data which are quite
sparse in the archives.  In active star--forming galaxies, where
UV--emission is dominated by massive stars, this thermal cold dust
($\approx\!30$\,K, Siebenmorgen et al. 1999\nocite{siekru_99}) mostly
peaks around the 50--100\,$\mu \rm m$ spectral region \cite[and
references therein]{cessau00}. As a consequence, the 60/100\,$\mu \rm
m$ fluxes ratio is very sensitive to the cold dust
temperature. Another reason why $\log(S_{60}/S_{100})$ may be chosen
as tracer of star formation is that, contrary to the mid-IR
($\approx\!10-40\,\mu m$), the transiently heated very small grains of
dust do not contaminate the 60--100\,$\mu m$ cold dust emission
\cite{cessau00}. Unfortunately the low angular resolution of IRAS data
($2\arcmin$ at $100\,\mu m$) only allows a global estimation of the
SFR intensity of a galaxy, regardless to its spacial distribution.

In view of the previous considerations, a link between Seyfert and/or
starburst activity and NIR colour profile can be probed in the diagram
$\Delta$(J--K') versus $\log(S_{60}/S_{100})$ plotted in
Fig.~\ref{fig3}. As already mentioned in Sect.~2.4, data from the
literature were added in order to obtain more reliable statistics. The
whole sample has Hubble Type ranging from $\approx\!-3$ up to
$\approx\!7$ (SAB0-SBc), and has its IR properties summarized in
Table~\ref{djmktab}.
%___________________________
\begin{table}[t]
\caption{Main properties of the sample plotted in Fig~\ref{fig3}.}
\label{djmktab}
\renewcommand{\thefootnote}{\alph{footnote}}
%\begin{tabular}{lcccc}
%\newcolumntype{Y}{>{\centering\arraybackslash}X}
%\newcolumntype{Z}{>{\small}Y}
%\begin{tabularx}{\linewidth}{YYYYYY}
%\begin{tabular}{@{}l@{\hspace{-3pt}}c@{\hspace{4pt}}c@{\hspace{5pt}}c@{
%\hspace{-3pt}}c@{\hspace{2pt}}c@{}}
\begin{tabular}{@{}l@{\hspace{-3pt}}c@{\hspace{2pt}}c@{\hspace{2pt}}c@{
\hspace{-3pt}}c@{\hspace{-1pt}}c@{}}\hline
Galaxies  &
${\rm log}(\frac{S_{60}}{S_{100}})$  &
$\Delta$(J--K')  &  
J--K'(1.5$\arcsec$) &
Activity&
Nested\\ 
\multicolumn{1}{c}{}		&
\multicolumn{1}{c}{}  &
\multicolumn{1}{c}{[mag]}       &
\multicolumn{1}{c}{[mag]}	&
\multicolumn{1}{c}{}		&
\multicolumn{1}{c}{Struct.}     \\
\multicolumn{1}{c}{(1)}          &
\multicolumn{1}{c}{(2)}  &
\multicolumn{1}{c}{(3)}          &
\multicolumn{1}{c}{(4)}	   &
\multicolumn{1}{c}{(5)}    &
\multicolumn{1}{c}{(6)}     \\
\hline
\object{NGC\,573}\footnotemark[1]	&$-$0.30 &   0.30  & 1.56 &Sy2 & B+B         \\
\object{NGC\,1068}\footnotemark[1]   	&$-$0.11 &   1.42  & 2.65 &STB/Sy2 & B+nS?       \\
\object{NGC\,2110}\footnotemark[1]    	&$-$0.15 &   0.62  & 2.03 &STB/Sy2 &             \\
\object{NGC\,2992}\footnotemark[1]    	&$-$0.32 &   0.36  & 1.94 &STB/Sy2 &             \\
\object{NGC\,3393}\footnotemark[1]    	&$-$0.22 &   0.14  & 1.09 &STB/Sy2 & B+B         \\
\object{NGC\,4253}\footnotemark[1]    	&$-$0.02 &   1.22  & 2.89 &STB/Sy1.5 &             \\
%\vspace{1truemm}
\object{NGC\,4388}\footnotemark[1]    	&$-$0.21 &   0.76  & 2.27 &STB/Sy2 &             \\
\\
\object{NGC\,470}\footnotemark[2]	&$-$0.28 &   1.41  & 1.47 &STB & B+T         \\
\object{NGC\,4314}\footnotemark[2]    	&$-$0.30 &   0.20  & 1.20 &STB/LIN & B+B         \\
\object{NGC\,6951}\footnotemark[2]    	&$-$0.44 &   0.14  & 1.23 &Sy2 & B+T         \\
\object{NGC\,7098}\footnotemark[2]    	&$-$0.67 &$-$0.09\ ~\ & 1.18 &  & B+T+B       \\
%\vspace{1truemm}
\object{NGC\,7479}\footnotemark[2]    	&$-$0.31 &   0.50  & 1.72 &STB/Sy2  & B+T         \\
\\
ESO\,215\footnotemark[3]
%-G031%
	&$-$0.24 &   0.34  & 1.26 &STB  & B+B         \\
ESO\,264\footnotemark[3]
%-G036%
	&$-$0.37 &   0.60  & 1.47 & &             \\
ESO\,320\footnotemark[3]
%-G030%
	&$-$0.11 &   0.55  & 1.34 &STB & B+B         \\
ESO\,323\footnotemark[3]
%-G077%
	&$-$0.21 &   0.64  & 1.85 &STB/Sy1 & B+T         \\
ESO\,374\footnotemark[3]
%-G032%
	& \ ~0.02 &   0.63  & 1.58 &STB & int.        \\
ESO\,443\footnotemark[3]
%-G017%
	&$-$0.13 &   0.38  & 1.34 &STB & B+B            \\
ESO\,508\footnotemark[3]
%-G005%
	&$-$0.40 &   0.20  & 1.13 &Sy2 & B+dB?       \\
NGC\,4903\footnotemark[3]    	&$-$0.39 &   0.23  & 1.17  &Sy2&             \\
NGC\,4939\footnotemark[3]    	&$-$0.59 &   0.25  & 1.42  &Sy2 &             \\
NGC\,4941\footnotemark[3]    	&$-$0.49 &   0.24  & 1.18  &Sy2 &             \\
NGC\,5135\footnotemark[3]    	&$-$0.28 &   0.15  & 1.22  &STB/Sy2& B+nS        \\
NGC\,5643\footnotemark[3]    	&$-$0.37 &   0.04  & 1.35  &Sy2&             \\
NGC\,6221\footnotemark[3]    	&$-$0.37 &   0.64  & 1.49  &Sy2& T+B            \\
NGC\,6300\footnotemark[3]    	&$-$0.46 &   0.81  & 1.79  &Sy2&             \\

\hline
\end{tabular}
\vspace{1truemm}

\footnotemark[1] Alonso--Herrero et al. \cite*{alosim_98}\\
\footnotemark[2] Paper II\\
\footnotemark[3] Our sample\\
(2): $S_{60}, S_{100}$ are non--colour corrected fluxes in the 60\,$\mu$m and 100\,$\mu$m
   IRAS--bands, from {\em Catalogued Galaxies + QSOs observed in IRAS
   Survey, Vers.2 (IPAC 1989)}\\
(3): defined in Sect. 3.2.2.\\
(4): J--K' integrated  within the fitted ellipse with a 1.5$\arcsec$ semi--major
   axis.\\
(5): type of activity; (Sy, LIN)$\,=\,$(Seyferts, LINER) from respective sources / 
STB$\,=\,$starbursts $\leftrightarrow \log(S_{60}/S_{100})\ga-0.35$. \\
(6): nested structures; symbols are defined in Table~3.
\end{table}

%**verif nS de 1068

%_________________________________________________________________

%{\bf
As mentionned above, $\Delta$(J--K') is independent of the photometric
calibrations. Thus its uncertainties mainly come from the readout
noise of the detector and are only marginal ($\leq 1\%$) compared to
the typical conservative error on $\log(S_{60}/S_{100})$ ($\approx 10-30
\%$). This later uncertainty is deduce from the errors on the individual 60 and 100 \,$\mu m$ IRAS fluxes given in the IRAS {\em Point Source Catalogue} (see e.g. Young et al. 86\nocite{yousch_86}) which are generally 5--15\%.
%} 

%In particular, because of the wide bandwidths of IRAS filters,
%the color correction could be as large as 50% (see The Explanatory
%Supplement ##). As this correction depends mainly on the assumed
%energy distribution of the source, we prefer not to apply it; that way
%we minimize random errors which are the more critical ones regarding
%to the detection of a correlation.
  
%The main source of uncertainty on the $\Delta$(J--K') comes from  
%Typical errors from the photometric calibration are  
%______
Independently of the presence of a Seyfert nucleus, the range of
$\Delta$(J--K') tends to increase from -0.1--0.9 up to 0.3--1.5 as
$\log(S_{60}/S_{100})$ increases.
%**DEBUT MODIF REFEREE...
%{\bf
Despite the efforts we made to extend the sample, it is still not
large enough to unambiguously point out a possible correlation in such a plot.
% But we have quantified the robustness of this tendancy by computing itscorrelation coefficient.
Using NIR colour profile tables given in Peletier et
al. \cite*{pelkna_99}, the same trend is observed for 29 objects among
their Seyfert sample.  As we do not have the original NIR data, these
new points are not reported in Fig.~\ref{fig3} and cannot be directly
compared to ours. But as they could only amplify the scatter of the
$\Delta$(J--K') vs $\log(S_{60}/S_{100})$ plot, they could be use to
quantify the robustness of the observed trend: while the slope and the
zero point of the regression remain nearly the same in both case,
the correlation coefficient is found to be 0.56 for the sample plot in
Fig.~\ref{fig3}, and 0.55 if the 29 Peletier et al. \cite*{pelkna_99}
objets are added. This low but nearly constant value of the
correlation coefficient, suggests that the link between
$\Delta$(J--K') and $\log(S_{60}/S_{100})$ is marginally linear but
real.
%...FIN MODIF**
%\bf}
 Moreover the
Fig.~\ref{fig3} looks very similar if one plots either the
earliest--types ($T\!<\!3$) or the latest--types ($T\!\ge\!3$), so
that it is not an Hubble sequence effect. Thus the {\em integrated}
FIR colour is related to the {\em nuclear} NIR colour (scaled to the
disc colour): significant starburst galaxies have central J--K'
$0.3-1.5$\,mag redder than the disc. Hunt et al. \cite*{hunmal_99}
found that Seyfert 1 and nuclear starburst galaxies have the bulge
J--K colour 0.1\,mag redder the disc, whereas Seyfert 2 bulges have
the same colour as the disc. Of course these results could not
directly be compared to ours but both are compatible: as the bulge
scale--length is always larger than the inner photometric aperture we
used, the contrast between their inner and outer colour naturally
tends to be lower than ours.

\subsection{Explanation attempt of the $\Delta$(J--K') excess}
The interpretation of the $\Delta$(J--K') red excess observed in
starburst galaxies is far from being straightforward. But the strong
correlation between $\log(S_{60}/S_{100})$ and the {\em nuclear}
H$\alpha$ luminosity density found among Markarian galaxies in
Mazzarella et al. \cite*{mazbot_91} might be a guideline and suggests
that, for their sample, the bulk of FIR luminosity arises from a
nuclear starburst (young enough to still ionize the surrounding
material). 

Could the red nuclear NIR colour we observed result from a recent
nuclear starburst?  Apart from the usual contribution of the old
stellar population, mainly giants, the J--K' colour depends at least
on three factors\footnote{the spectrum of the free--free emission
being almost flat in the NIR, its contribution to the J--K' colour
vanishes}: 1) the amount of 
% absorbent
dust (extinction is higher in J than in K'), 2) the gas and/or dust
temperature (James \& Seigar 1999\nocite{jamsei99} claimed that hot
dust contributes to K'), and 3) the respective contribution of various
types of young stars in star forming regions (K' luminosity increases
much more than J one if K supergiants dominate; J increases slightly
more than K' if OB stars dominate, James \& Seigar
1999\nocite{jamsei99}). Thus, in luminous and young star--forming
regions J--K' naturally tends to increase, so that the trend observed
in Fig.~\ref{fig3} may indicate that FIR luminosity is essentially
produced by a nuclear starburst.

None of the three factors having an effect on J--K' could alone be
responsible for the whole variations of $\Delta$(J--K') through our
sample. Indeed, using Leitherer et al. \cite*{leisch_99} starburst
model, we are unable to explain J--K' differences larger than
$\approx\!0.5$\,mag between an old stellar population and a recent
starburst. The maximum $\Delta$(J--K') we can obtain is
$\approx\!0.5-0.6$\,mag for a continuous SFR of 10\,Myr old with a
Salpeter IMF and a solar metallicity, and assuming a J-K'
$\!\approx\!0.5\,$mag for the old stellar population.

So additional factors that could also locally change J--K' have to be
considered, for instance: the 150--170\,K gas component associated with AGN and
starburst founded with ISO \cite{stulut_96} tends to increase
J--K'. The non--thermal continuum of PAHs or VSG likely contributes
more to the K'-band than the thermal emission of the 150--170\,K gas
\cite{routip_96}. The stellar and interstellar metallicity
gradient is another parameter, which effect on $\Delta$(J--K') could
not accurately be estimated without having previously disentangled the
age--dust degeneracy.

%% HW:PAS INCLU DANS LES MODELES STARBURST99 ??????
%The stellar and interstellar metallicity gradient is another important
%parameter
%(REF? review sur effet de metalicite sur NIR**) 
%which effect on $\Delta$(J--K') could not accurately be estimated
%without having previously disentangled the age--dust degeneracy. 

Thus, a general explanation about the trend observed in
Fig.~\ref{fig3} cannot be stated. It requires a careful study of the
gas and dust properties and of the stellar population in individual
objects, for which additional data are essential. For that purpose, H2
and Br$\gamma$ NIR narrow-band imaging of ESO\,215-G031 and NGC\,3081,
two double--barred galaxies, have already been performed and will be
used to carry on our investigations in a forthcoming paper.

\section{Conclusions}

This NIR study on 15 active (starburst and/or Seyfert)
objects leads us to these morphological results:
%detect these peculiar morphological features and :
\begin{enumerate}
\item[--]4 galaxies are double--barred systems: NGC\,3393 (already
  known), ESO\,215--G031, ESO\,320--G030, ESO\,443--G017 (new ones).
  These objects confirm previous studies which concluded that embedded
  bars are dynamically decoupled.
\item[--]4 galaxies have nested feature inside the primary bar: three
  present a triaxial bulge (ESO\,508--G005, ESO\,323--G077,
  NGC\,6221), one (NGC\,5135) harbours a nuclear spiral structure.
\item[--]We suggest to classify NGC\,5643 and ESO\,320--G030 as SB
  rather than SAB.
\item[--]Among the 2 unbarred galaxies (NGC\,4939, NGC\,4941), one
  (NGC\,4941) harbours a nuclear bar (like e.g. NGC\,7702, Paper I).
\item[--] The Seyfert or starburst activity is not directly linked to
  the presence of embedded structures: galaxies with embedded
  structures span a wide range of starburst activity and are found in
  Seyferts and non-Seyferts (see Fig.~\ref{fig3} and
  Table~\ref{djmktab}).
\end{enumerate}

A careful examination has been applied to $\mu_{J}-\mu_{K'}$, and
has lead us to the following considerations:
\begin{enumerate}
\item[--] The outer $\mu_{J}-\mu_{K'}$ profiles are roughly constant
  ($\approx\!0.8-1.2\,{\rm mag}\,\arcsec^{-2}$)
\item[--] For the whole sample, the $\mu_{J}-\mu_{K'}$ profiles
  increases toward the center or at least is constant at higher value
  than the disc.
%HW CETTE PHRASE N'EST PAS A PROPREMENT PARLER UNE CONCLUSION 
%HW \item[--] Seeing departure between J and K' can cause artificial
%HW $\mu_{J}-\mu_{K'}$ blue dip toward the nucleus.
%profiles leads to an inner J-K' colour redder than in the outer part.
\item[--] We have characterized these inner $\mu_{J}-\mu_{K'}$
profiles behaviour by the $\Delta$(J--K') parameter which seems to be:
\begin{enumerate} 
\item[i)] independent of the Seyfert activity,
\item[ii)] linked to the starburst indicator
$\log(S_{60}/S_{100})$.  Starburst galaxies tend to have a central
J--K' value 0.3--1.5\,mag higher than the disc one. This trend links
up the {\em nuclear} NIR colour to the {\em integrated} FIR colour.
\end{enumerate}
\item[--] A young nuclear starburst naturally tends to increase
  $\Delta$(J--K'), and could be partially responsible for the trend
  mentioned above.
%\item[--] Accurate dust maps have to be performed since classical
%  starburst models are enable to reproduce a
%  $\Delta$(J--K')$\!\ga\!0.5\,$mag, so that the dust may play a key
%  role. It is an indispensable prerequisite before studying the role
%  of other parameters on $\Delta$(J--K') (e.g. the metallicity or the
%  stellar population gradients).
\item[--] Classical starburst models are unable to reproduce a
  $\Delta$(J--K')$\!\ga\!0.5\,$mag, so that other factors or components
  have to be considered (e.g. dust extinction, the metallicity or the
  stellar population gradient, warm gaz).  Their quantitative incidence
  on $\Delta$(J--K'), requires additional data.
\end{enumerate}

%_________________________________________________________________
\begin{acknowledgements}
  This project is supported by the Swiss National Science Foundation
  (FNS). This research have made use of the NASA/IPAC Extragalactic
  Database (NED) which is operated by the Jet Propulsion Laboratory,
  California Institute of Technology, under contact with the U.S.
  Aeronautics and Space Administration. Partially based on
  observations made with the NASA/ESA Hubble Space Telescope, obtained
  from the data archive at the Space Telescope Science Institute.
  STScI is operated by the Association of Universities for Research in
  Astronomy, Inc. under NASA contract NAS 5-26555. We warmly thank
  Almudena Alonso--Herrero for its NIR data.
\end{acknowledgements}

\bibliography{aamnem99,eso59}

\begin{thebibliography}{}

\bibitem[\protect\astroncite{{Alonso-Herrero} et~al.}{1998}]{alosim_98}
{Alonso-Herrero}, A., {Simpson}, C., {Ward}, M.~J., and {Wilson}, A.~S., 1998,
\newblock {ApJ} {495}, 196

\bibitem[\protect\astroncite{{Buta}}{1995}]{but95}
{Buta}, R., 1995,
\newblock {ApJS} {96}, 39

\bibitem[\protect\astroncite{{Carter} and {Meadows}}{1995}]{carmea95}
{Carter}, B.~S. and {Meadows}, V.~S., 1995,
\newblock {MNRAS} {276}, 734

\bibitem[\protect\astroncite{{Cesarsky} and {Sauvage}}{2000}]{cessau00}
{Cesarsky}, C.~J. and {Sauvage}, M., 2000,
\newblock in D. {Block}, I. {Puerari}, A. {Stockton}, and W. {Ferreira }
  (eds.), {"Towards a New Millennium in Galaxy Morphology"}, Dordrecht,
  {Johannesburg, South Africa},
\newblock In press

\bibitem[\protect\astroncite{{Crocker} et~al.}{1996}]{crobau_96}
{Crocker}, D.~A., {Baugus}, P.~D., and {Buta}, R., 1996,
\newblock {ApJS} {105}, 353

\bibitem[\protect\astroncite{{de Vaucouleurs}}{1974}]{dev74}
{de Vaucouleurs}, G., 1974,
\newblock in {IAU Symp. 58: The Formation and Dynamics of Galaxies}, Vol.~58,
  p. 335

\bibitem[\protect\astroncite{{de Vaucouleurs} et~al.}{1991}]{rc3}
{de Vaucouleurs}, G., {de Vaucouleurs}, A., {Corwin}, J.~R., {Buta}, R.~J.,
  {Paturel}, G., and {Fouque}, P., 1991,
\newblock in {Third reference catalogue of bright galaxies (1991)}

\bibitem[\protect\astroncite{{D\'esert} et~al.}{1990}]{desbou_90}
{D\'esert}, F.~X., {Boulanger}, F., and {Puget}, J.~L., 1990,
\newblock {A\&A} {237}, 215

\bibitem[\protect\astroncite{{Dultzin-Hacyan} and {Benitez}}{1994}]{dulben94}
{Dultzin-Hacyan}, D. and {Benitez}, E., 1994,
\newblock {A\&A} {291}, 720

\bibitem[\protect\astroncite{{Forbes} and {Norris}}{1998}]{fornor98}
{Forbes}, D.~A. and {Norris}, R.~P., 1998,
\newblock {MNRAS} {300}, 757

\bibitem[\protect\astroncite{{Friedli}}{1999}]{fri99}
{Friedli}, D., 1999,
\newblock in J.~E. {Beckman} and T.~J. {Mahoney} (eds.), {ASP Conf. Ser.}

\bibitem[\protect\astroncite{{Friedli} and {Martinet}}{1993}]{frimar93}
{Friedli}, D. and {Martinet}, L., 1993,
\newblock {A\&A} {277}, 27

\bibitem[\protect\astroncite{{Friedli} et~al.}{1996}]{friwoz_96}
{Friedli}, D., {Wozniak}, H., {Rieke}, M., {Martinet}, L., and {Bratschi}, P.,
  1996,
\newblock {A\&AS} {118}, 461,
\newblock \bf Paper II \rm

\bibitem[\protect\astroncite{{Glass} and {Moorwood}}{1985}]{glamoo85}
{Glass}, I.~S. and {Moorwood}, A. F.~M., 1985,
\newblock {MNRAS} {214}, 429

\bibitem[\protect\astroncite{{Ho} et~al.}{1997}]{ho_fil_97}
{Ho}, L.~C., {Filippenko}, A.~V., and {Sargent}, W. L.~W., 1997,
\newblock {ApJ} {487}, 591

\bibitem[\protect\astroncite{{Hunt} and {Malkan}}{1999}]{hunmal99}
{Hunt}, L.~K. and {Malkan}, M.~A., 1999,
\newblock {ApJ} {516}, 660

\bibitem[\protect\astroncite{{Hunt} et~al.}{1999}]{hunmal_99}
{Hunt}, L.~K., {Malkan}, M.~A., {Moriondo}, G., and {Salvati}, M., 1999,
\newblock {ApJ} {510}, 637

\bibitem[\protect\astroncite{{Hunt} et~al.}{1997}]{hunmal_97}
{Hunt}, L.~K., {Malkan}, M.~A., {Salvati}, M., {Mandolesi}, N., {Palazzi}, E.,
  and {Wade}, R., 1997,
\newblock {ApJS} {108}, 229

\bibitem[\protect\astroncite{{James} and {Seigar}}{1999}]{jamsei99}
{James}, P.~A. and {Seigar}, M.~S., 1999,
\newblock {A\&A} {350}, 791

\bibitem[\protect\astroncite{{Jarvis} et~al.}{1988}]{jardub_88}
{Jarvis}, B.~J., {Dubath}, P., {Martinet}, L., and {Bacon}, R., 1988,
\newblock {A\&AS} {74}, 513

\bibitem[\protect\astroncite{{Jungwiert} et~al.}{1997}]{juncom_97}
{Jungwiert}, B., {Combes}, F., and {Axon}, D.~J., 1997,
\newblock {A\&AS} {125}, 479

\bibitem[\protect\astroncite{{Kaz\`es} et~al.}{1990}]{kazpro_90}
{Kaz\`es}, I., {Proust}, D., {Mirabel}, L.~F., {Combes}, F., {Balkowski}, C.,
  and {Martin}, J.~M., 1990,
\newblock {A\&A} {237}, L1

\bibitem[\protect\astroncite{{Knapen} et~al.}{2000}]{knashl_00}
{Knapen}, J.~H., {Shlosman}, I., and {Peletier}, R.~F., 2000,
\newblock {ApJ} {529}, 93

\bibitem[\protect\astroncite{{Koribalski}}{1996a}]{kor96a}
{Koribalski}, B., 1996a,
\newblock in {ASP Conf. Ser. 91: IAU Colloq. 157: Barred Galaxies}, p. 172

\bibitem[\protect\astroncite{{Koribalski}}{1996b}]{kor96b}
{Koribalski}, B., 1996b,
\newblock in {ASP Conf. Ser. 106: The Minnesota Lectures on Extragalactic
  Neutral Hydrogen}, p. 238

\bibitem[\protect\astroncite{{Leitherer} et~al.}{1999}]{leisch_99}
{Leitherer}, C., {Schaerer}, D., {Goldader}, J.~D., {Delgado}, R. M. G.~l.,
  {Robert}, C., {Kune}, D.~F., {de Mello}, D. l.~F., {Devost}, D., and
  {Heckman}, T.~M., 1999,
\newblock {ApJS} {123}, 3

\bibitem[\protect\astroncite{{Malkan} et~al.}{1998}]{malgor_98}
{Malkan}, M.~A., {Gorjian}, V., and {Tam}, R., 1998,
\newblock {ApJS} {117}, 25

\bibitem[\protect\astroncite{{M\'arquez} et~al.}{1999}]{mardur_99}
{M\'arquez}, I., {Durret}, F., {Delgado}, R. M. G.~l., {Marrero}, I.,
  {Masegosa}, J., {Maza}, J., {Moles}, M., {P\'erez}, E., and {Roth}, M., 1999,
\newblock {A\&AS} {140}, 1

\bibitem[\protect\astroncite{{Mazzarella} et~al.}{1991}]{mazbot_91}
{Mazzarella}, J.~M., {Bothun}, G.~D., and {Boroson}, T.~A., 1991,
\newblock {AJ} {101}, 2034

\bibitem[\protect\astroncite{{Mulchaey} and {Regan}}{1997}]{mulreg97}
{Mulchaey}, J.~S. and {Regan}, M.~W., 1997,
\newblock {ApJ Lett.} {482}, L135

\bibitem[\protect\astroncite{{Mulchaey} et~al.}{1997}]{mulreg_97}
{Mulchaey}, J.~S., {Regan}, M.~W., and {Kundu}, A., 1997,
\newblock {ApJS} {110}, 299

\bibitem[\protect\astroncite{{Peletier} et~al.}{1999}]{pelkna_99}
{Peletier}, R.~F., {Knapen}, J.~H., {Shlosman}, I., {P\'erez-Ram\'irez}, D.,
  {Nadeau}, D., {Doyon}, R., {Espinosa}, J. M.~R., and {Garc\'ia}, A. M.~P.,
  1999,
\newblock {ApJS} {125}, 363

\bibitem[\protect\astroncite{{Pfenniger}}{2000}]{pfe99}
{Pfenniger}, D., 2000,
\newblock in D. {Block}, I. {Puerari}, A. {Stockton}, and W. {Ferreira }
  (eds.), {"Towards a New Millennium in Galaxy Morphology"}, Dordrecht,
  {Johannesburg, South Africa},
\newblock In press

\bibitem[\protect\astroncite{{Quillen} et~al.}{1996}]{quiram_96}
{Quillen}, A.~C., {Ramirez}, S.~V., and {Frogel}, J.~A., 1996,
\newblock {ApJ} {470}, 790

\bibitem[\protect\astroncite{{Rouan} et~al.}{1996}]{routip_96}
{Rouan}, D., {Tiphene}, D., {Lacombe}, F., {Boulade}, O., {Clavel}, J.,
  {Gallais}, P., {Metcalfe}, L., {Pollock}, A., and {Siebenmorgen}, R., 1996,
\newblock {A\&A} {315}, L141

\bibitem[\protect\astroncite{{Rowan-Robinson} and {Crawford}}{1989}]{rowcra89}
{Rowan-Robinson}, M. and {Crawford}, J., 1989,
\newblock {MNRAS} {238}, 523

\bibitem[\protect\astroncite{{Sandage} and {Tammann}}{1981}]{santam81}
{Sandage}, A. and {Tammann}, G.~A., 1981,
\newblock {A revised Shapley-Ames Catalog of bright galaxies},
\newblock Washington: Carnegie Institution, 1981, Preliminary version

\bibitem[\protect\astroncite{{Shlosman} et~al.}{1989}]{shlfra_89}
{Shlosman}, I., {Frank}, J., and {Begelman}, M.~C., 1989,
\newblock {Nat} {338}, 45

\bibitem[\protect\astroncite{{Siebenmorgen} et~al.}{1999}]{siekru_99}
{Siebenmorgen}, R., {Kr\"ugel}, E., and {Chini}, R., 1999,
\newblock {A\&A} {351}, 495

\bibitem[\protect\astroncite{{Sturm} et~al.}{1996}]{stulut_96}
{Sturm}, E., {L\"utz}, D., {Genzel}, R., {Sternberg}, A., {Egami}, E., {Kunze},
  D., {Rigopoulou}, D., {Bauer}, O.~H., {Feuchtgruber}, H., {Moorwood}, A.
  F.~M., and {de Graauw}, T., 1996,
\newblock {A\&A} {315}, L133

\bibitem[\protect\astroncite{{Tsvetanov} and {Petrosian}}{1995}]{tsvpet95}
{Tsvetanov}, Z.~I. and {Petrosian}, A.~R., 1995,
\newblock {ApJS} {101}, 287

\bibitem[\protect\astroncite{{Vega Beltr\'an} et~al.}{1998}]{vegzei_98}
{Vega Beltr\'an}, J.~C., {Zeilinger}, W.~W., {Amico}, P., {Schultheis}, M.,
  {Corsini}, E.~M., {Funes}, J.~G., {Beckman}, J., and {Bertola}, F., 1998,
\newblock {A\&AS} {131}, 105

\bibitem[\protect\astroncite{{V\'eron-Cetty} and {V\'eron}}{1993}]{verver93}
{V\'eron-Cetty}, M.~P. and {V\'eron}, P., 1993,
\newblock {A Catalogue of quasars and active nuclei},
\newblock ESO Scientific Report, Garching: European Southern Observatory (ESO),
  |c1993, 6th ed.

\bibitem[\protect\astroncite{{Wainscoat} and {Cowie}}{1992}]{waicow92}
{Wainscoat}, R.~J. and {Cowie}, L.~L., 1992,
\newblock {AJ} {103}, 332

\bibitem[\protect\astroncite{{Wozniak} et~al.}{1995}]{wozfri_95}
{Wozniak}, H., {Friedli}, D., {Martinet}, L., {Martin}, P., and {Bratschi}, P.,
  1995,
\newblock {A\&AS} {111}, 115,
\newblock \bf Paper I \rm

\bibitem[\protect\astroncite{{Young} et~al.}{1986}]{yousch_86}
{Young}, J.~S., {Schloerb}, F.~P., {Kenney}, J.~D., and {Lord}, S.~D., 1986,
\newblock {ApJ} {304}, 443

\end{thebibliography}
\bibliographystyle{aabib99}

%_________________________________________________________________
%-> exemple d'inclusion de figure avec "graphics":
%
%\begin{figure*}[p]
%\resizebox{\hsize}{!}{\includegraphics{toto.ps}}
%\caption{...}
%\label{fig:toto}
%\end{figure*}

\setcounter{figure}{1}
\begin{figure*}
%\infig{8.8}{art_prof_e215.ps}{17.6}
%\unitlength=1truemm
%\begin{picture}(200,290)(20,-20)
%\put(0,0){\epsfig{file=test.ps,width=200mm,height=290mm}}
%\end{picture}
%\resizebox{210mm}{!}{\includegraphics{test.ps}}
%\epsfig{file=test.ps,width=200mm,height=200mm}
%\infig{22}{figure3.ps}{20}
\caption[]{For each of the fifteen galaxies, five (or six) panels are shown: 
  {\bf Top left.}  Radial profiles of surface brightness $\mu$,
  ellipticity $e$, and position angle PA. J--band is on the left,
  K'--band on the right.  Note the square root scale.  {\bf Top
    right.} J and K' grey--scale images with contours superimposed.
  The spacing is 0.4\,mag. North is up, East on the left. {\bf Bottom
    left.} J--K' radial profile. {\bf Bottom right.} HST NICMOS
  F160W frame (when available) and J--K' colour map}
\label{figprofil}
\end{figure*}

\end{document}